     \documentstyle[12pt]{article}
     \newlength{\dinwidth}
     \newlength{\dinmargin}
\def\Journal#1#2#3#4{{#1} {\bf #2}, #3 (#4)}

\def\NPB{{\em Nucl. Phys.} B}
\def\PLB{{\em Phys. Lett.}  B}
\def\PRL{\em Phys. Rev. Lett.}

\def\ZPC{{\em Z. Phys.} C}
\def\lsim{\mathrel{\rlap{\lower4pt\hbox{\hskip1pt$\sim$}}
    \raise1pt\hbox{$<$}}}                
\def\gsim{\mathrel{\rlap{\lower4pt\hbox{\hskip1pt$\sim$}}
    \raise1pt\hbox{$>$}}}                
\begin{document}
\input{epsf}
\vspace*{10mm}
\begin{center}  \begin{Large} \begin{bf}
RADGEN 1.0 \\ Monte Carlo Generator for Radiative Events
in \\ DIS on Polarized and Unpolarized Targets\\
  \end{bf}  \end{Large}
  \vspace*{5mm}
  \begin{large}
I.Akushevich$^a$, H.B\"ottcher$^b$, D.Ryckbosch$^c$\\
  \end{large}
\end{center}
$^a$ NC PHEP, Bogdanovich str. 153, 220040 Minsk, Belarus\\
$^b$ DESY Zeuthen, 15738 Zeuthen, Germany\\
$^c$ University of Gent, 9000 Gent, Belgium
\begin{quotation}
\noindent
{\bf Abstract:} A new Monte-Carlo
generator including real radiated photons in DIS on polarized and
unpolarized targets is
presented.
Analytical and numerical tests are performed
and discussed in details.
\end{quotation}
\section {Introduction }

Experimental data on lepton-nucleon scattering  contain contributions 
from the usual 
Born process and from QED radiative effects which come
from loop corrections and from processes with the emission of
additional real photons.  Any Monte-Carlo program simulating the
experimental situation has to take into account these radiative
processes by generating radiated photons 
because radiative corrections can be very
large.

Depending on the four-momentum transfer squared ($Q^2$) and the
energy transfer ($\nu$) there are three basic channels for lepton
scattering on nuclei, namely elastic, quasielastic, and inelastic
processes.  In the case of elastic scattering ($\nu=Q^2/2M_A$,
where $M_A$ is nuclear mass) the leptons are scattered off the
nucleus leaving the nucleus in its ground state.  Quasielastic
scattering ($\nu\sim Q^2/2M$, where $M$ is nucleon mass)
corresponds roughly to direct collisions with the individual
nucleons inside the nuclei.  The inelastic scattering occurs when
the pion threshold is reached ($\nu \ge Q^2/2M+m_{\pi}$ where
$m_\pi$ is the pion mass).  At the Born level both $Q^2$ and $\nu$ are
fixed completely by measuring the scattering angle and the energy
(momentum) of the scattered lepton.  
  However, at the level of radiative corrections, in case of presence of
real radiated photons, the fixation is removed and the 
four-momentum of the radiated photon has to be included in the kinematical
variable calculation.
 Such elastic,
quasielastic and inelastic processes with radiation of a real
photon are known as radiative tails from the elastic
($\sigma_{el}$) and the quasielastic ($\sigma_{q}$) peaks and from
the continuous spectrum ($\sigma_{in}$) called hereafter shortly
elastic, quasielastic and inelastic radiative tail.

\begin{figure}[t]
\begin{tabular}{ccccc}
\begin{picture}(60,100)
\put(30,60){\line(2,-1){20.}}
\put(50,50){\line(2,1){20.}}
\put(50,17.5){\circle*{10.}}
\put(55,16.5){\line(2,-1){15.}}
\put(54.4,14){\line(2,-1){15.6}}
\put(54.4,20){\line(2,-1){15.6}}
\multiput(50,28)(0,8){3}{\oval(4.0,4.0)[r]}
\multiput(50,24)(0,8){4}{\oval(4.0,4.0)[l]}
\put(30,10){\line(2,1){15.}}
\put(30,9){\line(2,1){15.}}
\put(50,-10){\makebox(0,0){\small a)}}
\end{picture}
&
\begin{picture}(60,100)
\multiput(40,57)(0,8){3}{\oval(4.0,4.0)[r]}
\multiput(40,61)(0,8){3}{\oval(4.0,4.0)[l]}
\put(30,60){\line(2,-1){20.}}
\put(50,50){\line(2,1){20.}}
\put(50,17.5){\circle*{10.}}
\put(55,16.5){\line(2,-1){15.}}
\put(54.4,14){\line(2,-1){15.6}}
\put(54.4,20){\line(2,-1){15.6}}
\multiput(50,28)(0,8){3}{\oval(4.0,4.0)[r]}
\multiput(50,24)(0,8){4}{\oval(4.0,4.0)[l]}
\put(30,10){\line(2,1){15.}}
\put(30,9){\line(2,1){15.}}
\put(50,-10){\makebox(0,0){\small b)}}
\end{picture}
&
\begin{picture}(60,100)
\multiput(60,57)(0,8){3}{\oval(4.0,4.0)[r]}
\multiput(60,61)(0,8){3}{\oval(4.0,4.0)[l]}
\put(30,60){\line(2,-1){20.}}
\put(50,50){\line(2,1){20.}}
\put(50,17.5){\circle*{10.}}
\put(55,16.5){\line(2,-1){15.}}
\put(54.4,14){\line(2,-1){15.6}}
\put(54.4,20){\line(2,-1){15.6}}
\multiput(50,28)(0,8){3}{\oval(4.0,4.0)[r]}
\multiput(50,24)(0,8){4}{\oval(4.0,4.0)[l]}
\put(30,10){\line(2,1){15.}}
\put(30,9){\line(2,1){15.}}
\put(50,-10){\makebox(0,0){\small c)}}
\end{picture}
&
\begin{picture}(60,100)
\put(30,60){\line(2,-1){20.}}
\put(50,50){\line(2,1){20.}}
\put(50,17.5){\circle*{10.}}
\put(55,16.5){\line(2,-1){15.}}
\put(54.4,14){\line(2,-1){15.6}}
\put(54.4,20){\line(2,-1){15.6}}
\multiput(50,28)(0,8){3}{\oval(4.0,4.0)[r]}
\multiput(50,24)(0,8){4}{\oval(4.0,4.0)[l]}
\multiput(42,55)(4,4){3}{\oval(4.0,4.0)[lt]}
\multiput(42,59)(4,4){2}{\oval(4.0,4.0)[br]}
\multiput(50,63)(4,-4){3}{\oval(4.0,4.0)[tr]}
\multiput(54,63)(4,-4){2}{\oval(4.0,4.0)[bl]}
\put(30,10){\line(2,1){15.}}
\put(30,9){\line(2,1){15.}}
\put(50,-10){\makebox(0,0){\small d)}}
\end{picture}
&
\begin{picture}(60,100)
\put(30,60){\line(2,-1){20.}}
\put(50,50){\line(2,1){20.}}
\put(50,17.5){\circle*{10.}}
\put(55,16.5){\line(2,-1){15.}}
\put(54.4,14){\line(2,-1){15.6}}
\put(54.4,20){\line(2,-1){15.6}}
\multiput(50,48)(0,8){1}{\oval(4.0,4.0)[r]}
\multiput(50,44)(0,8){1}{\oval(4.0,4.0)[l]}
\multiput(50,28)(0,8){1}{\oval(4.0,4.0)[r]}
\multiput(50,24)(0,8){1}{\oval(4.0,4.0)[l]}
\put(50,36){\circle{12.}}
\put(30,10){\line(2,1){15.}}
\put(30,9){\line(2,1){15.}}
\put(50,-10){\makebox(0,0){\small e)}}
\end{picture}
\end{tabular}
\vspace{0.5cm}
\caption{
Feynman diagrams contributing to the Born and the radiative correction
cross sections in lepton-nucleus scattering.
}
\label{feyn}
\end{figure}
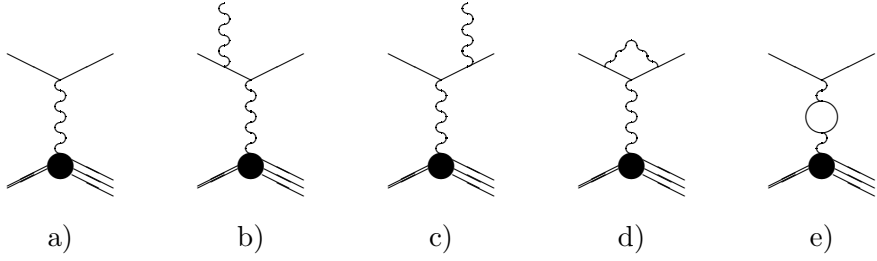

The total radiative correction at the lowest order is obtained as the
sum of these contributions together with loop corrections ($\sigma_v$)
coming from effects of vacuum polarization and exchange of an additional
virtual photon:

\begin{equation}
\sigma^{rad.corr.}=\sigma_{in}+\sigma_{q}+\sigma_{el}+\sigma_{v}.
\label{rc00}\end{equation}

In this report we present a Monte-Carlo generator for
events with a possible radiation of a real photon in deep inelastic
lepton scattering (DIS) on polarized and unpolarized targets. 
  It should be noted that only the single photon exchange (no $Z^0$
exchange) and only pure QED corrections are considered. Hence the use of
the generator is at present restricted to experiments where the
electroweak contributions and corrections are small, as e.g. in the
experiment HERMES at DESY.
The
events are generated in accordance with their contribution to the
observed total cross section given by 
\begin{equation}
\sigma_{obs}=\sigma_{non-rad}(\Delta)+\sigma_{in}(\Delta)
+\sigma_{q}+\sigma_{el}.
\label{addrc00}\end{equation}
The first term $\sigma_{non-rad}(\Delta)$ 
contains not only the contribution from the
Born process but also the contributions from loop corrections
($\sigma_v$) and from multiple soft photon production with a total energy
not exceeding a cut-off parameter $\Delta$. In handling the radiative
corrections two approaches have been used, the one developed by Mo and
Tsai \cite{MT} and the one given by Bardin and Shumeiko \cite{BSh}. 
These two 
formalisms and the significance of the cut-off parameter $\Delta$ will
be discussed below in more detail. 
The generator contains two patches which
can be considered as independent generators.
The first one is based on the FORTRAN code POLRAD 2.0 \cite{POLRAD20}
handling the polarized lepton-nucleon scattering. The other is based on
FERRAD \cite{FERRAD}, a code that deals with the unpolarized case. They
will be referred to 
as the POLRAD
and FERRAD generator throughout the text.  A detailed
analytical and numerical comparison of the results given by the
two generators is performed.

\section {The Kinematics and Main Stages of Generation}

The Feynman diagram for the Born (or one photon exchange) process
is presented in Fig.\ref{feyn}a. Its kinematics 
is completely defined by the scattering angle $\theta$
and the energy $E'$ of the scattered lepton, two variables
which are usually measured. 
All other inclusive
kinematical variables can be expressed in terms of $\theta$ and
$E'$. Having in mind the laboratory frame, i.e. the frame with the target
nucleus at rest the relevant variables are 
\begin{equation}\begin{array}{c}\displaystyle
Q^2=4EE'\sin^2\frac{\theta}{2}, \qquad
\nu=E-E', \qquad
x={Q^2\over 2M\nu},
\\[0.5cm]\displaystyle
y=\frac{\nu}{E},\qquad
W^2=Q^2(1/x-1)+M^2,
\end{array}\label{0010}\end{equation}
where $E$ is the beam energy, $Q^2$ is the negative of the 
four-momentum transfer squared, $\nu$ the
energy transfer, $x$ the Bjorken scaling variable, $y$ the normalized
energy transfer, and $W^2$ the mass squared of the hadronic final state.
Radiative QED corrections (RC) at the lowest order are described by
the set of the Feynman graphs shown in fig. \ref{feyn}b-e. Contributions
to the lowest order RC cross section come from the diagrams b) and c)
and from the interference of the loop diagrams d) and e) with the Born
matrix element a).

An event registered in the detector with certain values of
$E'$ and $\theta$ for the scattered lepton can either be a
non-radiative  or a radiative event, i.e. an event containing a real hard 
radiated photon.
For a radiative event there are apart from $E'$ and
$\theta$ three additional variables necessary to fix the kinematics of
the radiated real photon. A possible choice is the
photon energy $E_{\gamma}$ and the two angles $\theta_{\gamma}$ and
$\phi_{\gamma}$, where $\theta_{\gamma}$
is the angle between the real and the virtual photon momenta
$\vec k$ and $\vec q=\vec k_1-\vec k_2$ ($k_{1,2}$ being the
momenta of the initial and the scattered lepton) and
$\phi_{\gamma}$ is the angle between the planes defined by the
momenta
($\vec k_1$, $\vec k_2$) and ($\vec k$, $\vec q$). For events with
the radiation of a real photon the kinematical variables describing the
virtual photon and used to generate the hadronic final
state differ from eq.(\ref{0010}) because the substitution  $q \rightarrow q-k$ has
to be
made in their definition. The variables obtained after this
 substitution will be referred to as 'true' ones:
\begin{equation}
\begin{array}{ll}\displaystyle
W^2_{true}=W^2-2E_{\gamma}(\nu +
M-\sqrt{\nu^2+Q^2}\cos\theta_{\gamma}),
&\displaystyle
\nu_{true}=\nu-E_{\gamma},
\\[0.5cm] \displaystyle
Q^2_{true}=Q^2+2E_{\gamma}(\nu
-\sqrt{\nu^2+Q^2}\cos\theta_{\gamma}),
&\displaystyle
x_{true}={Q^2_{true} \over 2M\nu_{true}}.
\end{array}
\label{true}
\end{equation}
For non-radiative events the true kinematics exactly coincide with
(\ref{0010}).

In case of the elastic and quasielastic radiative processes the mass
squared of the hadronic final state is fixed imposing additional
constraints on the true kinematical variables. Indeed the following
ranges are allowed:
\begin{equation}
\left\{
\begin{array}{cl}
\displaystyle x\leq x_{true}\leq 1 & \displaystyle{\rm for}\;\;
\sigma_{in}\\[0.5cm]
\displaystyle x_{true}= 1 & \displaystyle{\rm for}\;\;
\sigma_{q}\\[0.5cm]
\displaystyle x_{true}=M_A/M & \displaystyle{\rm for}\;\;
\sigma_{el}
\end{array}
\qquad
Q^2_{min}\leq Q^2_{true}\leq Q^2_{max}
\right.
\label{0011}\end{equation}
where
\begin{equation}
Q^2_{max,min}
=Q^2{2(1-x_r)(1\pm \sqrt{1+\gamma^2})+\gamma^2 \over \gamma^2+4x_r(1-x_r)}
\label{0012}\end{equation}
and
\begin{equation}
\gamma^2=4M^2x^2/Q^2, \qquad x_r=x/x_{true}.
\label{0013}\end{equation}

The Monte-Carlo procedure for the generation of events with a possible
photon radiation
is as follows:

\noindent
It starts off with the generation of the kinematics of the scattered
lepton and the calculation of an event weight from these kinematics.
Then the appropriate scattering channel (non-radiative; elastic,
quasielastic or inelastic radiative tail) has to be chosen according
to their contribution to the total observed cross section (see
(\ref{addrc00})).
If a radiative channel is selected the radiated photon has to be
generated and the values of the kinematical variables have to be
re-computed in order to obtain the true values. For each event the
weight has to be recalculated. The new weight is defined as the
ratio of the radiatively corrected and the Born cross section. After
this recalculation the weighted sum of all events (generated originally
in accordance with the Born cross section) gives the observed cross
section.

The recalculation of the weight requires the knowledge of the cross
section integrated over the photon momentum. This is done differently
in the two theoretical approaches which will be discussed in detail in
the subsections below. In the Mo-Tsai approach an additional parameter
$\Delta$ is introduced dividing the integration region into a soft and a
hard photonic part. No such parameter is used in the Bardin-Shumeiko
approach.
However, in both approaches a minimal photon energy $ E_{min}^{\gamma}$
is adapted in generating a radiated photon. Only if the photon energy is
above this value the kinematics of the photon is calculated and the event
becomes a radiative one. The actual value of $ E_{min}^{\gamma}$ depends
on the aim of the photon generator. In general photons should be detected
in the calorimeter. So the energy threshold of the calorimeter sets the
value for $ E_{min}^{\gamma}$.

In addition one needs spin averaged and spin dependent structure
functions, quasielastic suppression factors and elastic formfactors
to calculate the Born and observed cross sections. This is taken either
from fits to
experimental data or from model predictions. A recent 8-parameter fit
from ref.\cite{nmc} for the proton and deuterium structure function
$F_2^{p,d}(x,Q^2)$ and their combination for
$^3$He ($F_2^{^3{\rm He}}=\frac{1}{3}F_2^p+\frac{2}{3}F_2^d$) has
been used. 
The spin dependent structure function  $g_1(x,Q^2)$ is constructed
from a fit to the spin asymmetry $A_1$ \cite{POLRAD20} according to
\begin{equation}
g_1(x,Q^2)=A_1(x)F_1(x,Q^2)=
{A_1(x)F_2(x,Q^2)\over
2x\biggl(1+R(x,Q^2)\biggr)}\bigl(1+\gamma^2\bigr).
\label{0001}\end{equation}

For the structure function $R(x,Q^2)$ being the ratio of longitudinal to
transverse virtual photon absorption cross sections Whitlow's fit
\cite{Whi} was used and assumed to be A-independent. No contributions are
assumed from the spin dependent structure function $g_2(x,Q^2)$. For the
elastic formfactors a fit to experimental data \cite{ffhe} is utilized.
The quasielastic radiative tail is calculated in accordance with
ref.\cite{ARST}. In contrast to the elastic case where the treatment is
exact (within the considered order of perturbative theory) here the
Y-scaling hypothesis \cite{Tho} and the peaking approximation
(quasielastic structure functions are evaluated at the quasielastic peak)
are applied.

The two Monte Carlo generators for radiative events available in
RADGEN
for the unpolarized and polarized cases are described below. The
theoretical approaches, the approximations made, and the
dependence of the radiative corrections on the artificial parameter
$\Delta$ and on the value of the photon detector threshold
$ E^{\gamma}_{min}$ are discussed.

\subsection{The generator for the unpolarized case}

The unpolarized generator is based on the FORTRAN code FERRAD35
\cite{FERRAD}. The code
calculates the radiative correction to deep inelastic scattering
of unpolarized particles in accordance with the analytical formulae
given by Mo and Tsai \cite{MT}.

The result for the lowest order radiative correction of the cross
section is

\begin{equation}
\sigma  = \delta_{R}(\Delta)(1+\delta_{vert}+\delta_{vac}+\delta_{sm})\sigma_{1\gamma}
+ \sigma_{el}+ \sigma_{q}+ \sigma_{in}(\Delta)
,
\label{001}\end{equation}

\noindent
where $\sigma_{1\gamma}$ is the one photon exchange Born cross section and
$\delta_{vac}$, $\delta_{vert}$, and $\delta_{sm}$ are
corrections due to vacuum polarization by electron and muon pairs, 
vertex corrections and
residuum of the cancellation of infrared divergent terms independent of
$\Delta$:
\begin{eqnarray}\label{0033}
\delta_{vac}&=&\frac{\alpha}{\pi}\left[
-\frac{20}{9}+\frac{2}{3}\ln\frac{Q^2}{m_e^2}
+\frac{2}{3}\ln\frac{Q^2}{m_{\mu}^2}\right]
,\nonumber\\
\delta_{vert}&=&\frac{\alpha}{\pi}\left[
-2+\frac{3}{2}\ln\frac{Q^2}{m^2} \right]
,\nonumber\\
\delta_{sm}&=&\frac{\alpha}{\pi}\left[
-\frac{\pi^2}{6}+{\rm Li}_2\biggl(cos^2\frac{\theta}{2}\biggr)
-\frac{1}{2}\ln^2(1-y)
\right],
\end{eqnarray}
where $m$ is the lepton mass and 
${\rm Li}_2(x)=-\int_0^xdy \ln(1-y)/y$ is the dilogarithm or
Spence function.

The cross sections $\sigma_{el}$, $\sigma_{q}$, and $\sigma_{in}$
are the contributions from radiative processes (i.e. including real photon
radiation) for
elastic, quasielastic and deep inelastic scattering. They can be
calculated in terms of the radiative tail from the j$th$ mass level
$\sigma_j$

\begin{equation}
\sigma_{el}=\sigma_{el}(M_A,1), \qquad
\sigma_q=\sigma_q(M,1), \qquad
\sigma_{in}=\int\limits_{M+m_{\pi}}^WdM_h\sigma_{in}(M_h,\theta_{max}),
\label{a00a}\end{equation}
where $M_h=\sqrt{W^2_{true}}$ and $\sigma_j$ ($j=el,q,in$) has the form of
an integral
over $\theta_{\gamma}$, the angle between the real and the virtual photons:

\begin{equation}
\sigma_j(M,\theta_{max})
=\int\limits_{0}^{\theta_{max}}
d\theta_{\gamma}\;\; T_0(W_2^j(T_1+T_2+T_3+T_4+T_5)+W_1^j(T_6+T_7)).
\label{a008}\end{equation}

The explicit form of the structure functions $W_{1,2}^j$ depends on the
type of the tail. The terms $T_i$ are kinematical factors (see 
ref.\cite{MT} for details). The integration limit $\theta_{max}$ is
defined as $\theta_{max}=min(\pi,\theta_{\Delta})$ where $\theta_{\Delta}$
can be found from the relation

\begin{equation}
W^2_{true}=W^2-2\Delta\bigl(\nu +
M-\sqrt{\nu^2+Q^2}\cos(\theta_{\Delta})\bigr).
\label{a004}\end{equation}

An artificial parameter $\Delta$ had to be introduced to divide the
integration region over the photon energy into two parts, the soft and the
hard energy regions. The hard energy region $\sigma_{in}(\Delta)$ can be
calculated without any approximations. The soft photon part is calculated
for photon energies approaching zero. After cancellation of the infrared
divergences and a resummation of soft multiphoton effects the correction
factor is given by 
\begin{equation} 
\delta_R(\Delta)=\exp\left[-
\frac{\alpha}{\pi}\left(\ln \frac{E}{\Delta}+\ln \frac{E'}{\Delta}\right)
\left(\ln \frac{Q^2}{m^2}-1\right)\right]. 
\label{002}\end{equation} 

Note that we follow ref.\cite{MT} in fixing that part of the soft photon
correction which should be exponentiated.
 Another possible variant is
the well-known prescription of ref.\cite{Suura}. The dependence on $\Delta$
is the deficiency of the method.
On the one hand this artificial parameter has to be chosen sufficiently
large
to avoid numerical instabilities because for $\Delta=0$ the
contribution $\sigma_{in}(\Delta)$ gets infinite.
On the other hand it has to be chosen sufficiently small to reduce the
soft photon region which is calculated only approximately.
If the threshold over the photon energy is low enough, it is customary to
use the same value for the
parameters $\Delta$ and $ E^{\gamma}_{min}$.

\subsection{The generator for the polarized case}

The polarized generator is constructed utilizing the FORTRAN code
POLRAD 2.0 \cite{POLRAD20,ASh}. This code is based on the method
of covariant cancellation of infrared divergences developed by Bardin
and Shumeiko
 \cite{BSh}. The final formula for the cross section free of infrared
divergences does not include any artificial parameter like
$\Delta$. However, $ E^{\gamma}_{min}$ has to be introduced if the code
is used considering the generation of real radiated photons. The cross
section formula of ref.\cite{ASh} can be rewritten into a form similar
to eq.(\ref{001}):

\begin{equation}
\sigma  = \delta_{R}(
E^{\gamma}_{min})(1+\delta_{vert}+\delta_{vac}+\delta_{sm})\sigma_{1\gamma}
+\sigma_{add}( E^{\gamma}_{min})
+ \sigma_{el}+ \sigma_{q}+ \sigma_{in}( E^{\gamma}_{min}).
\label{007}\end{equation}

In the ultrarelativistic approximation ($Q^2\gg m^2$) the
corrections $\delta_{vert}$, $\delta_{vac}$, and $\delta_{sm}$ take the
same form as in eq.(\ref{0033}) \footnote{It should be noted that
contributions from $\tau$-leptons and quark loops were included into
$\protect\delta_{vac}$ \protect\cite{ASh}.}. The covariant form of the
radiative tails for a nuclear target with atomic number $A$ is as follows:
\begin{eqnarray}\label{b001}
\sigma _{in}&=& - {\alpha ^{3}y}\int\limits^{\tau_{max}}_{\tau_{min}}
d\tau\sum^{4}_{i=1}
 \sum^{k_{i}}_{j=1} \theta _{ij}(\tau)\int\limits^{R_{max}}
 _{R_{min}} dR
 {R^{j-2}\over (Q^2+R\tau)^{2}}\Im_{i}(R,\tau),
\nonumber\\
\sigma _{el}&=&
- {\alpha ^{3}y\over A^2}\int\limits^{\tau_{Amax}}_{\tau_{Amin}}d\tau_A
 \sum^{4}_{i=1}\sum^{k_{i}}_{j=1} \theta _{ij}(\tau_A){
 2M^{2}_A R^{j-2}_{el\;A}\over
(1+\tau_A)(Q^2+R_{el\;A}\tau_A)^{2}}\Im
^{el}_{i}(R_{el},\tau_A),
\nonumber\\
\sigma _{q}&=&
- {\alpha ^{3}y\over A}\int\limits^{\tau_{max}}_{\tau_{min}}d\tau
 \sum^{4}_{i=1}\sum^{k_{i}}_{j=1} \theta _{ij}(\tau){
 2M^{2} R^{j-2}_{el}\over (1+\tau)(Q^2+R_{el}\tau)^{2}}\Im
^{q}_{i}(R_{q},\tau).
\end{eqnarray}
The integration limits are determined by
$\tau_{max,min}=Q^2/2M^2x(1\pm\sqrt{1+4M^2x^2/Q^2})$,
$R_{min}=2M E^{\gamma }_{min}$, and
$R_{max}=(W^2-(M+m_{\pi})^2)/(1+\tau )$.
The quantities $\theta _{ij}$ are kinematical factors, and $\Im_i$ are
some combinations of the DIS structure functions, elastic formfactors or
quasielastic response functions (see 
ref.\cite{ASh} for details).
The sum over $i$ corresponds to contributions from spin averaged
($i=1,2$) and spin dependent ($i=3,4$) structure functions.
In the case of the elastic and quasielastic radiative tail
there is no integration over $R$ because the $R$ range is reduced to
the constant values $R=R_{el\;A}$ and $R=R_{el}$
($R_{el}=(2M\nu-Q^2)/(1+\tau)$) due to the fixed final hadron
mass. Invariants with the index "A" contain the nucleus momentum $p_A$
instead of $p$ ($p_A^2=M_A^2$).

Apart from the usual ultrarelativistic approximation another one was made
when 
the eqs. (\ref{001}) and (\ref{002}) were obtained, namely the photon
energy was considered to be small $ E^{\gamma} \ll E,E'$ in the region
$E^{\gamma}<
E^{\gamma}_{min}$. 
In POLRAD, the term $\sigma_{add}(E_{min}^{\gamma})$ 
is added to take the difference of this approximation to the exact
formula
into account:
\begin{eqnarray}\label{010}
\sigma_{add}( E^{\gamma}_{min}) &=& -{\alpha ^{3}y}
 \int\limits^{\tau_{max}}_{\tau_{min}}d\tau \sum^{4}_{i=1}
 \left\{ \theta _{i1}(\tau)
\int\limits^{2M E^{\gamma}_{min}}_{0}{dR\over R} \right.
\left[ {\Im _{i}(R,\tau )\over (Q^2+R\tau )^2}-{\Im
_{i}(0,0)\over Q^4}\right]
\nonumber\\
&&\qquad \qquad\qquad\qquad\qquad\left.
+ \sum^{k_{i}}_{j=2} \theta _{ij}(\tau)\int\limits^{2M
E^{\gamma}_{min}}_{0}
dR {R^{j-2}\over (Q^2+R\tau )^{2}}\Im _{i}(R,\tau)  \right\}.
\end{eqnarray}
Therefore, the results obtained with the help of POLRAD are
independent of the threshold parameter $E_{min}^{\gamma}$ by
construction.

\section{Comparison and Numerical Results}

\subsection{Analytical comparison}

The expressions for the spin independent part of the radiative tails
in eq. (\ref{b001}) have been compared analytically with the radiative
tails given by eqs.(\ref{a00a}) and (\ref{a008}) for the unpolarized case
using the algebraic programming system REDUCE 3.5. It turned out that

\begin{eqnarray}\label{011}
-2M\sqrt{\nu^2+Q^2} \sum_{j=1}^3\Theta_{2i}(\tau)R^{j-3} &=&
T_1+T_2+T_3+T_4+T_5,
\nonumber\\
-2M\sqrt{\nu^2+Q^2} \sum_{j=1}^3\Theta_{1i}(\tau)R^{j-3} &=& T_6+T_7.
\end{eqnarray}

From these relations one can easily show that the formulae for the
radiative tails are exactly the same when no spin is considered.

All corresponding terms in (\ref{001}) and (\ref{007}) coincide exactly and
there is only one additional term $\sigma_{add}$ in eq. (\ref{007}).
This term can be considered as characterizing the quality of
            the soft photon approximation and can
be calculated analytically by expanding
it in the parameter $\epsilon=2M E^{\gamma}_{min}/Q^2$:

\begin{equation}
\sigma_{add}( E^{\gamma}_{min})
=\epsilon \frac{\alpha}{\pi}{\bar \sigma_0}-\epsilon\frac{\alpha^3y}{Q^2}
(\delta_{\tau}F_1+\delta_{22}F_2) +O(\epsilon^2),
\label{e001}\end{equation}
where ${\bar \sigma_0}$ is obtained in terms of the Born cross section
$\sigma_0$:

\begin{equation}
{\bar \sigma_0}=
 \sigma_0 \left\{
\begin{array}{l}
\displaystyle
F_1\rightarrow J_0x^2\frac{\partial F_1}{\partial x}
-\delta_{\tau}\biggl(-2F_1+x\frac{\partial F_1}{\partial x}
+Q^2\frac{\partial F_1}{\partial Q^2}
\biggr)
\\[0.3cm]
\displaystyle
F_2\rightarrow J_0\biggl(x^2\frac{\partial F_1}{\partial x}+xF_2\biggr)
-\delta_{\tau}\biggl(-2F_2+x\frac{\partial F_2}{\partial x}
+Q^2\frac{\partial F_2}{\partial Q^2}
\biggr).
\end{array}
\right.
\label{e002}\end{equation}
The quantities $J_0$, $\delta_{\tau}$, and $\delta_{22}$ depend only on
kinematical variables:

\begin{equation}\begin{array}{c}\displaystyle
\delta_{\tau}=\frac{xy}{1-y}\biggl(1-2\ln \frac{2E'}{m}\biggr)
-xy\biggl(1-2\ln \frac{2E}{m}\biggr), \qquad
J_0=2(\ln\frac{Q^2}{m^2}-1),
\\[0.5cm]\displaystyle
\delta_{22}=-4\frac{2-y}{y}\ln(1-y)-2J_0-\frac{4M^2x}{Q^2}\delta_{\tau}.
\end{array}\label{e003}\end{equation}
The first term in eq.(\ref{e001}) corresponds to the approximation
$E_{\gamma}=0$ in arguments of structure functions (terms with
derivatives)
and the photonic propagator (terms without derivative). The second
term in eq.(\ref{e001}) appears when the approximation is applied to
kinematical terms like $T_{1..7}$ as in eq.(\ref{a008}).

The numerical analysis performed for
HERMES kinematics shows that the difference between both generators
in the unpolarized case due to this additional contribution is
negligible.

\begin{figure}[t]
\unitlength 1mm
\begin{picture}(160,130)
\put(10,-10){
\epsfxsize=15cm
\epsfysize=15cm
\epsfbox{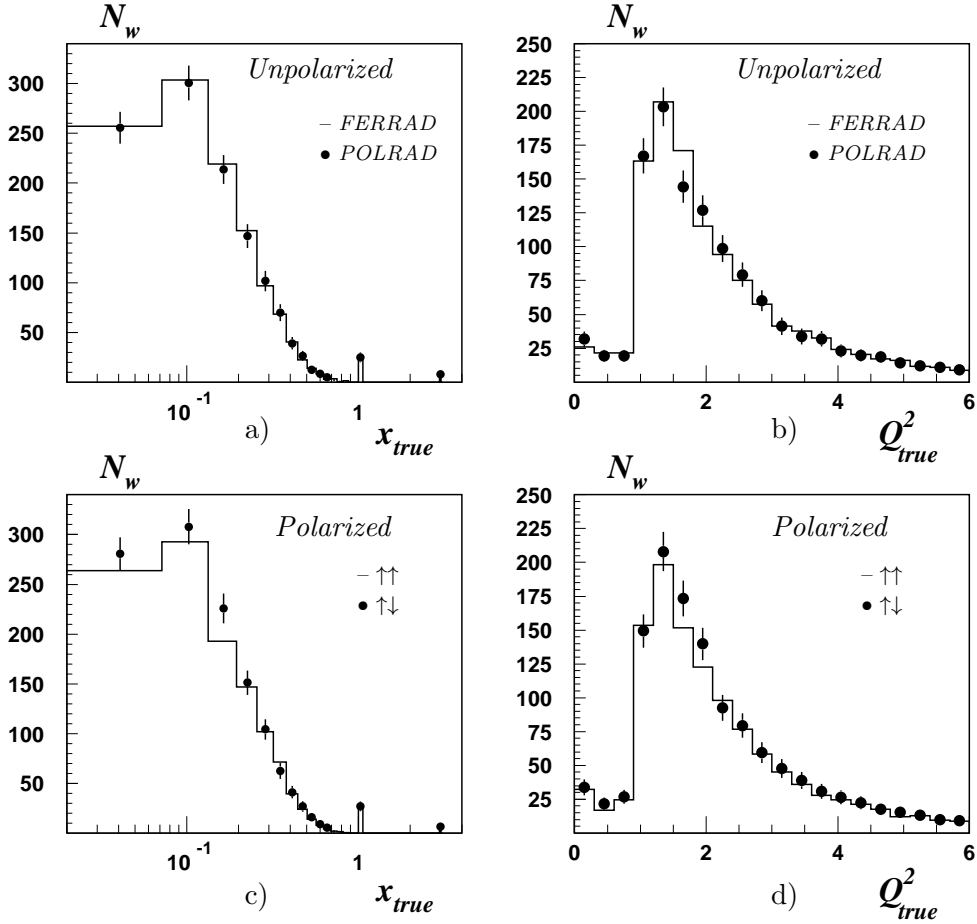}
}
\put(50,113){{\rm \small {\em Unpolarized}}}
\put(60,106){{\rm \scriptsize -- {\em FERRAD}}}
\put(60,102){{\rm \scriptsize $\bullet$ {\em POLRAD}}}
\put(115,113){{\rm \small {\em Unpolarized}}}
\put(125,106){{\rm \scriptsize -- {\em FERRAD}}}
\put(125,102){{\rm \scriptsize $\bullet$ {\em POLRAD}}}
\put(54,52){{\rm \small {\em Polarized}}}
\put(132,46){{\rm \scriptsize --  $\uparrow \uparrow$}}
\put(132,42){{\rm \scriptsize $\bullet$ $\uparrow \downarrow$}}
\put(120,52){{\rm \small {\em Polarized}}}
\put(65,46){{\rm \scriptsize --  $\uparrow \uparrow$}}
\put(65,42){{\rm \scriptsize $\bullet$ $\uparrow \downarrow$}}
\put(50,65){{\rm \small a)}}
\put(120,65){{\rm \small b)}}
\put(50,3){{\rm \small c)}}
\put(120,3){{\rm \small d)}}
\end{picture}
\caption{\protect\it \label{distr}
The weighted $x_{true}$ and $Q^2_{true}$ distributions 
as generated by FERRAD (histogram) and
POLRAD with polarization part switched off (points) a), b); and by POLRAD
with parallel (histogram) and antiparallel (points) spin configurations c), d).
}
\end{figure}


\begin{figure}[t]
\unitlength 1mm
\begin{picture}(160,130)
\put(10,-10){
\epsfxsize=15cm
\epsfysize=15cm
\epsfbox{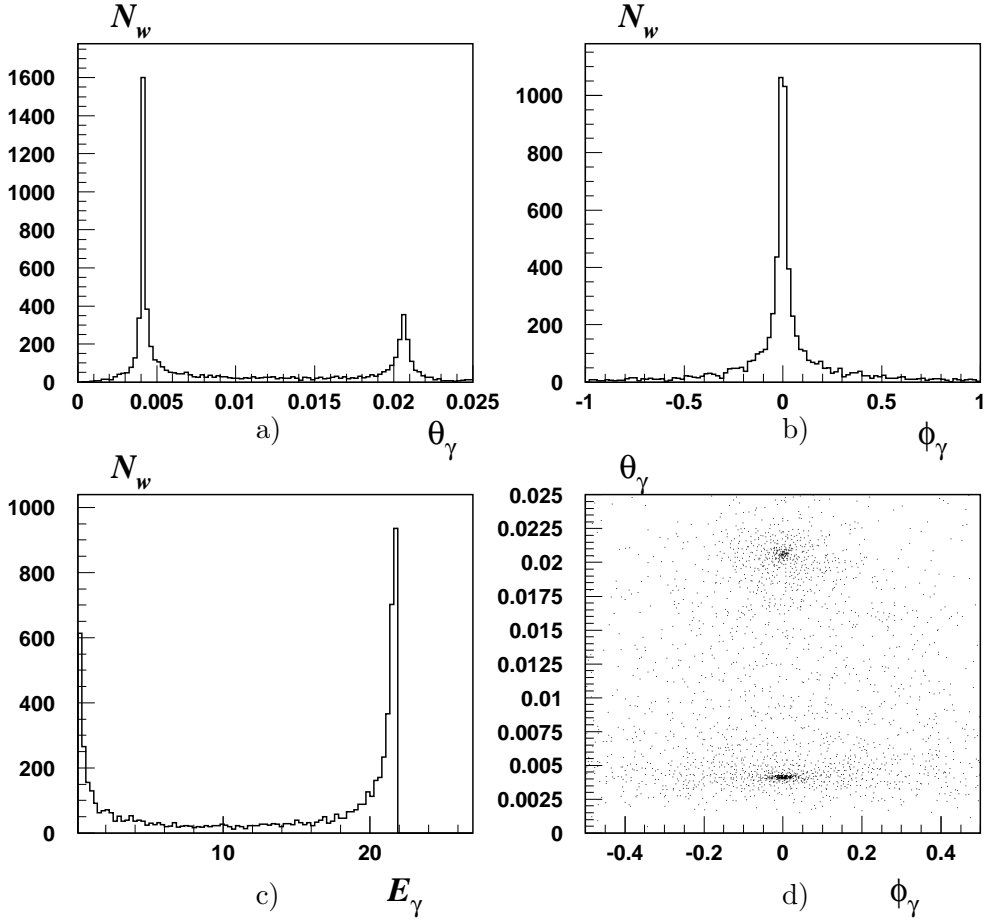}
}
\put(50,65){{\rm \small a)}}
\put(120,65){{\rm \small b)}}
\put(50,3){{\rm \small c)}}
\put(120,3){{\rm \small d)}}
\end{picture}
\caption{\protect\it \label{disg}
The distribution of the radiation angles $\theta_{\gamma}$ a) and
$\phi_{\gamma}$ b) and of the energy c) of the
radiated photon for $x=0.1$ and $y=0.8$. The two-dimensional
distribution d) shows $\theta_{\gamma}$ vs $\phi_{\gamma}$.}
\end{figure}

\begin{figure}[t]\centering
\unitlength 1mm
\parbox{.4\textwidth}{\centering
\begin{picture}(80,80)
\put(-5,0){
\epsfxsize=8cm
\epsfysize=8cm
\epsfbox{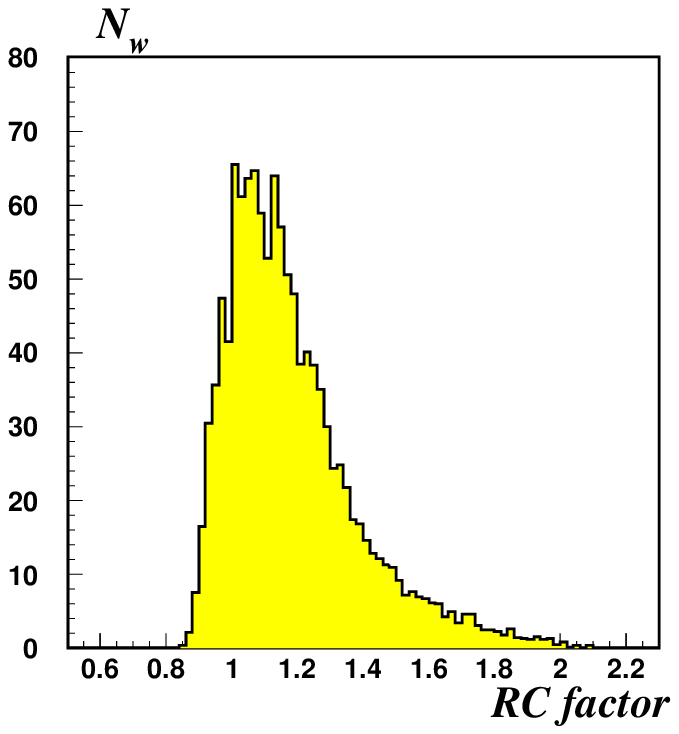}
}
\end{picture}
\caption{\label{sigcor}   
The typical distribution of the radiative 
correction factor.} 
}
\hfill
\parbox{.4\textwidth}{\centering
\begin{picture}(80,80)
\put(-8,0){
\epsfxsize=8cm
\epsfysize=8cm
\epsfbox{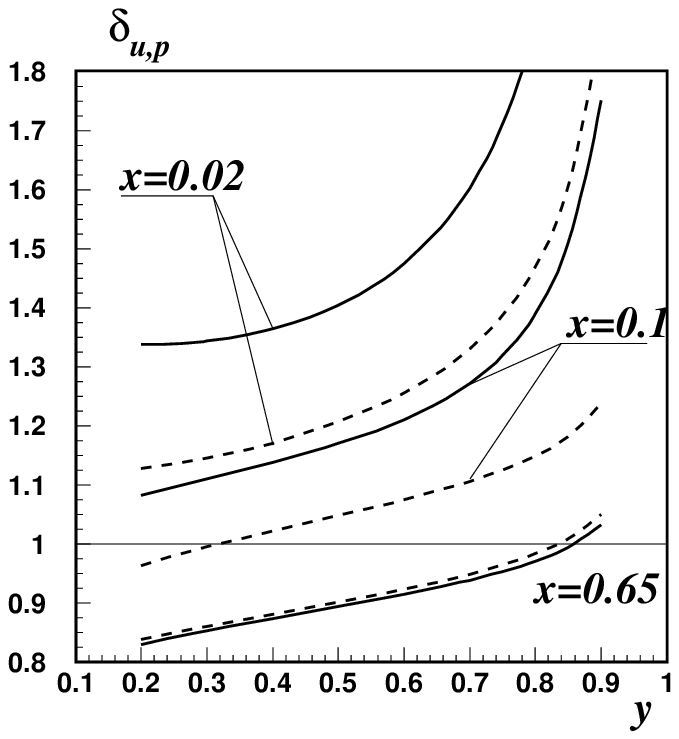}
}
\end{picture}
\caption{\label{allu} The spin averaged (full lines) and spin dependent
(dashed 
lines) RC factor as a function of $y$ for for different values of $x$.
}}
\end{figure}

\subsection {Numerical comparison}


In order to compare these two generators numerically Monte-Carlo event
samples of 100k events each have been generated for a $^3$He target and kinematical cuts relevant for the
HERMES experiment \cite{HERMES} have been applied 
($Q^2>$1 GeV$^2$, $W^2>$4 GeV$^2$, $y<$0.85, 0.037$<\theta<$0.14$rad$).  
Weighted distributions ($N_w$) of $x_{true}$ and $Q^2_{true}$ generated by the
FERRAD and the POLRAD generators are shown 
in Fig.\ref{distr}. In Figs. \ref{distr}a and \ref{distr}b
the unpolarized case is compared, i.e. FERRAD
and POLRAD with the polarization part switched off. As can be seen the
distributions are in agreement.
It should be noted that the events at $x\approx 1$ and $x\approx 3$
correspond to the quasielastic and elastic radiative tails,
respectively. 
In Fig. \ref{distr}c) and \ref{distr}d) a comparison of
the parallel and antiparallel spin configuration in POLRAD is
given. Here a systematic difference is seen which is expected from a
spin asymmetry different from zero at small values of $x$.

The distributions of the radiation angles
$\theta_{\gamma}$ and $\phi_{\gamma}$ as defined in 
the so-called Tsai-system in which the $z$-axis is along the
direction of the virtual photon and the $y$-axis is normal to the
scattering plane (see ref.\cite{MT}) and of the energy
$E_{\gamma}$ are displayed in Fig.\ref{disg}. These distributions are
generated for a certain point in the kinematical plane of the scattered
lepton, i.e. for $x$ = 0.1 and $y$ = 0.8.
The two peaks seen in the $\theta_{\gamma}$ distribution of Fig.\ref{disg}
correspond to the $s$- and $p$-peak emerging from collinear radiation
along the direction of the incoming and outgoing lepton, respectively.


Further results on the numerical comparison of the polarized and
unpolarized case in the covariant and Mo-Tsai approach, respectively, can
be found in the refs.\cite{KSh,BarBad}. Fig.4 of ref.\ \cite{BarBad}
presents the $\Delta$-dependence of the radiative correction factor
obtained by the code FERRAD. The general conclusion was that the best
choice is $\Delta$ equal to about 0.1\% of the beam energy. In this
case the results should not depend on $\Delta$.

The other task of the generators is to generate a radiative
correction factor $\sigma_{obs}/\sigma_{Born}$ necessary to
recalculate the weight of each event. A typical distribution of the RC
factor is shown 
in Fig.\ref{sigcor}. Apart from this integration characteristics it is
useful to consider also the behavior of the RC factor as a function of
kinematical variables. Both the observed and Born cross sections can be
split into spin averaged and spin dependent parts
($\sigma^{obs,Born}=\sigma^{obs,Born}_u + p_bp_t\sigma^{obs,Born}_p$,
$p_{b,t}$
are beam and target polarizations) 
allowing to define the RC factors for the unpolarized and
polarized case separately:  
\begin{equation} 
\delta_{u,p}=\frac{\sigma^{obs}_{u,p}}{\sigma^{Born}_{u,p}}.
\end{equation}
Here we keep in mind that there is a region ($x\sim$0.4 for $^3$He case)
where
$\sigma^{Born}_{p}$=0. The $y$-dependence of these factors for several
values of $x$ is plotted in Fig.\ref{allu}. We would like to mention that
the RC to the
 polarization
asymmetry $A_{obs,Born}=\sigma_p^{obs,Born}/\sigma_u^{obs,Born}$ can be written
 in terms of
$\delta_{u,p}$ as
\begin{equation}
{ A_{obs}-A_{Born} \over A_{Born} }= {\delta_p-\delta_u\over \delta_u}
\label{asy}\end{equation}
As  can be seen from Fig.\ref{allu} the correction factors $\delta_p$ and
$\delta_u$ 
have similar behaviour and become almost equal with increasing $x$.
The large 
corrections coming from the factorized part of eqs.(\ref{001}) and
(\ref{007}) (the first term) cancel exactly in the numerator of 
(\ref{asy}) and only the
unfactorized part of the corrections coming from the radiative tails
contribute to the RC of the polarization asymmetry.
 This is
the reason why there is a large correction to the total cross section and only a
small one to the asymmetry.

\section {Conclusion}

The Monte-Carlo generator RADGEN 1.0 including radiative events in DIS
on polarized and unpolarized targets have been
presented. Two patches of the generator, one for the unpolarized
case based on the FORTRAN code
FERRAD35 and the other for the polarized case based on the FORTRAN
code POLRAD 2.0 were compared and tested. This
generator is implemented into the HERMES
Monte-Carlo program HMC. The FORTRAN code of the
Monte Carlo generator {RADGEN 1.0}
is available on
request.

\section*{Acknowledgements}

We are grateful to N.Shumeiko for help and support.  We thank L.Favart
and H.Spiesberger for their thoughtful comments that helped clarify the
discussion in this paper. We would also like to thank N.Akopov, D.Bardin,
N.Gagunashvili, P.Kuzhir, A.Nagaitsev for useful discussions and comments.

\appendix
\renewcommand{\thesection}{Appendix:}
\renewcommand{\theequation}{\Alph{section}.\arabic{equation}}

\section{\bf Main options and keys}
\setcounter{equation}{0}

In this Appendix we briefly discuss options and keys which should
be specified by a user in order to run RADGEN. They all are
gathered in one input file 'input.dat' (default values are given
in brackets):
\begin{itemize}
\item {\bf nevent} [1000] -- number of events;
\item {\bf ixytest} [0] -- usage of look-up table for channel
generation. The information about results of RC calculation in the
50$\times$50 points over $x$ and $y$ is stored in this
look-up table. The key defines the way how to use this
information:

\begin{tabular}{cl}
 1 & two-dimensional spline\\
 2 &FINT interpolation\\
 0 &generation of the look-up table\\
 -1 & No use of the  look-up table
\end{tabular}
\item {\bf ige} [2] -- unpolarized ({1}) or polarized ({2})
generator is
used;
\item {\bf ire} [1] -- type of target: proton ({ 1}), deuteron
({2}), helium-3
({3});
\item {\bf e1} [27.5 GeV] -- beam energy;
\item {\bf pl} and {\bf pn} [0 and 0] -- degrees of beam and
target polarization;
\item {\bf ikey} [0] -- type of generation of leptonic kinematics
\begin{description}
\item[{\bf 1} -- ] {\bf x} and {\bf y} are generated in
accordance of the
cross section within kinematical cuts,
\item[{ \bf 0} --] { \bf x} and { \bf y} are read from the
standard input;
\end{description}
\item { \bf ntk} [35] -- 
number of bins in each of the seven parts of the integrand
over the azimuthal photonic angle. 
The region over the azimuthal photonic angle is divided into seven
parts in order to locate the peaks. This key defines the number of
bins in each part of the integrand over the azimuthal photonic angle
where the cross section is calculated.
\item {\bf nrr}
[200] -- number of bins in the integrand over
photonic
energy. This key defines the number of the bins over photon energy
where cross section of the process is calculated and stored.
\item { \bf demin} [0.1 GeV] -- { $\Delta$} the threshold of
photonic energy. Events
with photonic energy smaller  than{ $\Delta$} are considered as
non-radiative.
\item Kinematical cuts on $ y$, $ x$, $ Q^2$, $ W^2$,
$ \theta$.
\end{itemize}

\end{document}